\documentclass[twoside]{article}
\usepackage{qic}
\usepackage{amsmath}

\newcommand{\dsone}{\text{\usefont{U}{dsrom}{m}{n}1}}

\def\bra#1{\mathinner{\langle{#1}|}}
\def\ket#1{\mathinner{|{#1}\rangle}}
\def\braket#1#2{\langle{#1}|#2\rangle}

\begin{document}
\setlength{\textheight}{8.0truein}    

\runninghead{Generating quantum dissonance via local operations}
            {G. Torun}

\normalsize\textlineskip
\thispagestyle{empty}
\setcounter{page}{1}


\vspace*{0.25truein}

\alphfootnote

\fpage{1}

\centerline{\bf
GENERATING QUANTUM DISSONANCE}
\vspace*{0.035truein}
\centerline{\bf VIA LOCAL OPERATIONS}
\vspace*{0.37truein}
\centerline{\footnotesize
G\"{O}KHAN TORUN\footnote{Corresponding author: gokhan.torun@alanyauniversity.edu.tr; gtorung@gmail.com}}
\vspace*{0.015truein}
\centerline{\footnotesize\it Department of Computer Engineering, Alanya University,}
\baselineskip=10pt
\centerline{\footnotesize\it 07400 Alanya, Antalya, T\"{u}rkiye}
\baselineskip=10pt
\vspace*{10pt}

\vspace*{0.21truein}

\abstracts{
Correlations may arise in quantum systems through various means, of which the most remarkable one is quantum entanglement. Additionally, there are systems that exhibit non-classical correlations even in the absence of entanglement. Quantum dissonance refers to how quantum discord (QD) --- the difference between the total correlation and the classical correlation in a given quantum state --- appears as a non-classical correlation in a system without entanglement. It could be said that QD has the potential to provide a more inclusive viewpoint for discerning the non-classical correlations. In this work, we address the problem of manipulating the QD between two subsystems through local operations. We propose two explicit procedures for obtaining separable Werner states, a type of mixed state with nonzero QD. Both approaches involve performing local operations on classically correlated states and offers a step-by-step method for obtaining separable Werner states with nonzero discord, providing an alternative (explicit and user-friendly) to existing methods.
}{}{}

\vspace*{10pt}

\vspace*{3pt}

\vspace*{1pt}\textlineskip    

\section{Introduction}

From the smallest subatomic particles to the largest structures, correlations play a vital role in our understanding of the natural world. In a more specific context, quantum correlations (QCs) are of paramount importance as they provide a greater insight into the foundational concepts of quantum mechanics \cite{GAdesso2017-BookQD}. More importantly, these correlations have significant applications \cite{Adesso-2016-QCorrelations} in quantum information science and, therefore, enable the development of quantum technologies due to their non-classical nature, which provides unique advantages not achievable with classical resources. It is then important to certify QCs \cite{Modi2010UnifiedView, Xu2010ExpClassQua, Jia2020HierarchyofGQCs, Pusuluk2022OrbitalCorr} to continue advancing quantum technologies. Understanding QCs --- how they work, how to measure them, and how to manipulate them --- is especially crucial \cite{GAdesso2017-BookQD, Adesso-2016-QCorrelations, Lami2018Nonclassical, Jirakov2021ExpHieQCs, Demirel2021DiscordQComputaion, Liu2022CharacCorr, Louren2022GenuineCorr, Ganardi2022hierarchyof} for exploring the differences between classical and quantum physics \cite{Modi2012Review}.

Due to its ability to quantify the non-classical nature of correlations in a more general way, quantum discord (QD) \cite{Zurek2001-QD, Henderson2001-QCC} has become a valuable tool for studying the unique properties of quantum systems and their potential applications \cite{GAdesso2017-BookQD, Adesso-2016-QCorrelations}. In a nutshell, QD is a measure of the QC present in a quantum system, which extends beyond entanglement, as it demonstrates non-classical correlations even in the absence of entanglement \cite{Zurek2001-QD, Henderson2001-QCC}. In addition and importantly, it has been experimentally shown that separable states with nonzero QD, referred to as quantum dissonance, may have high probability of success in various quantum information processing tasks \cite{Datta2008QDis, Lanyon2008QDis}. In recent years, therefore, QD has gained significant attention in the field of quantum information science, due to its potential applications in tasks such as quantum communication \cite{Gao2019DiscordQCommunication}, state discrimination \cite{Roa2011DissStateDis, Li2012DissAssistedStateDis, Pang2013SeqStateDis, Zhang2013DissStateDis, Siyouri2022QDissonance}, quantum state merging \cite{Madhok2011StateMerging, Cavalcanti2011StateMerging}, remote state preparation \cite{Dakic2012RSP}, and quantum cryptography \cite{Pirandola2014QDResource}. In light of these studies, it becomes evident that there is a noteworthy aspect to QD that warrants deeper investigation.

Generally speaking, the examination of quantum features and correlations as valuable resources constitutes one of the primary objectives within the context of general quantum resource theories (RTs) \cite{Chitambar2019-QRTs, Kuroiwa2020GenQRTs, Torun2023MAJQRTs, Gour2024resources}. However, based on the current state of the literature, no RT framework specifically addresses QD. In fact, the lack of a shared consensus regarding the resource content of QD stands out as a major impediment to the formulation of a RT framework for QD. For instance, the absence of an established RT for QD can be attributed to its distinct response to local operations, where it can increase, setting it apart from other QCs such as quantum entanglement \cite{Horodecki2009ReviewQEnt}. Establishing a dedicated RT for quantum discord would not only bridge this gap but also contribute to a deeper comprehension of the quantification, manipulation, and potential applications of quantum correlations. Accordingly, within the context of establishing a RT for discord, studies examining the evolution of discord under local operations retain essential significance.

In this work, we focus on obtaining separable Werner state(s) characterized by nonzero quantum dissonance --- QD in the absence of entanglement. These states, which may have a high probability of success in remote state preparation \cite{Dakic2012RSP}, are generated through local operations starting from a classically correlated state, that is, a zero discord state. In Section \ref{Sec:Preliminary}, we cover some basic concepts. These include classically correlated states (Section \ref{Subsec:ClassicallyCorrelated}), which form the initial step of the problem under discussion; the rank of the correlation matrix (Section \ref{Subsec:CorrelationRank}), crucial for determining the rank of the initial state; QD (Section \ref{Subsec:QuantumDiscord}), which serves as the primary focus of our investigation; and the set of nonzero discord states (Section \ref{Subsec:Set-of-Nonzero-QD}), it is essential to clarify the necessary and sufficient conditions for nonzero discord. In Section \ref{Sec:Results}, we present two methods for transforming a given classically correlated state into another one with a nonzero discord value. First, Werner states (Section \ref{Subsec:WernerStates}) are discussed as the target states. Second, we discuss the generation of Werner states through local operations (Section \ref{Subsec:GeneratingNonzeroWerner}), presenting two accessible methods for achieving this transformation. In Section \ref{Sec:Conclusion}, we conclude our work with a summary. Given QD's direct relationship with the nonorthogonality of basis states, we address some potentially useful points in this regard.


\section{The Essential Phases of Our Approach}\label{Sec:Preliminary} 

In this section, we go over some definitions required to carry out our step-by-step approach. This sets the stage for our discussions regarding the generation of discord (more precisely, quantum dissonance) within an easily followed structure.


\subsection{Classically correlated states}\label{Subsec:ClassicallyCorrelated}

We begin with a (bipartite) composite quantum system described by \(\mathcal{H}_{AB}=\mathcal{H}_A\otimes\mathcal{H}_B\), which represents the Hilbert space resulting from the tensor product of the Hilbert spaces \(\mathcal{H}_A\) and \(\mathcal{H}_B\) corresponding to the subsystems controlled by Alice (\(A\)) and Bob (\(B\)), respectively. Then, a state \(\rho_{AB} \in \mathcal{H}_{AB}\) that is shared between \(A\) and \(B\) is called separable (or unentangled) if it has a decomposition in the form \cite{Werner1989EPRCorr}
\begin{eqnarray}\label{SEP-RhoAB}
\rho_{AB} = \sum_{i}p_i\rho_A^{(i)}\otimes\rho_B^{(i)}.
\end{eqnarray}
Here, \(\{\rho_A^{(i)}\}\) and \(\{\rho_B^{(i)}\}\) represent states associated with subsystems \(A\) and \(B\), respectively, and \(\{p_{i} \geq 0\}\) form a probability distribution, subject to the constraint \(\sum_{i}p_{i}=1\). Importantly, a given state \(\rho_{AB}\) is entangled (or nonseparable) if it cannot be written in the form of Eq.~(\ref{SEP-RhoAB}).

Moreover, the state represented in Eq.~(\ref{SEP-RhoAB}) can be further classified. Namely, the state given by Eq.~(\ref{SEP-RhoAB}) is classified as classical-quantum (CQ) if it can be expressed as:
\begin{eqnarray}\label{Class-Quant}
\rho_{AB}=\sum_{i}p_{i}\ket{\alpha_i}\bra{\alpha_i}_A\otimes\rho_B^{(i)},
\end{eqnarray}
where \(\{\ket{\alpha_i}_{A}\}\) form an orthonormal basis on Alice's Hilbert space \(\mathcal{H}_A\) and \(\{p_{i}\geq0\}\) with \(\sum_{i}p_{i}=1\). Hence, one can characterize the state given by Eq.~(\ref{Class-Quant}) as being classically correlated with subsystem \(A\) \cite{Adesso-2016-QCorrelations, Piani2012ClassCorre}. Likewise, when one classifies based on subsystem \(B\) and includes orthonormal basis states \(\{\ket{\beta_i}_{B}\}\), the resulting state becomes quantum-classical (QC) written \(\rho_{AB}=\sum_{i}p_{i}\rho_A^{(i)} \otimes \ket{\beta_i}\bra{\beta_i}_B\). Correspondingly, a state is classified as classical-classical (CC) if it can be expressed as:
\begin{eqnarray}\label{Class-Class}
\rho_{AB}=\sum_{i,j}p_{ij}\ket{\alpha_i}\bra{\alpha_i}_A\otimes\ket{\beta_j}\bra{\beta_j}_B.
\end{eqnarray}
In this context, the set \(\{\ket{\alpha_i}_{A}\}\) constitutes an orthonormal basis in Alice's Hilbert space \(\mathcal{H}_A\), while the set \(\{\ket{\beta_j}_{B}\}\) forms an orthonormal basis on Bob's Hilbert space \(\mathcal{H}_B\). Additionally, the coefficients \(\{p_{ij}\geq0\}\) are constrained by the usual condition \(\sum_{i,j}p_{ij}=1\). The initial step in addressing the problem under investigation involves commencing with the CC states as presented in Eq.~\eqref{Class-Class}, which is also called fully classically correlated.


\subsection{Rank of correlation matrix}\label{Subsec:CorrelationRank}

The definition of the rank of correlation matrix introduced by  Daki\'{c} \emph{et al}. \cite{Dakic2020NonzeroQDCondition} is particularly important for the problem  in which we are interested, as the rank of the classically correlated state at hand is determined accordingly. As expounded in Ref.~\cite{Dakic2020NonzeroQDCondition}, the initial step involves the selection of two basis sets within the local Hilbert-Schmidt spaces of Hermitian operators, denoted as \(\{A_n: n=1,2,\dots,d_A^2\}\) and \(\{B_m: m=1,2,\dots,d_B^2\}\), where \(d_A\) and \(d_B\) represent the dimensions of the respective local Hilbert-Schmidt spaces \(\mathcal{H}_A\) and \(\mathcal{H}_B\). Then, a given state \(\rho\) in \(\mathcal{H}_A \otimes \mathcal{H}_B\) can be represented as
\begin{eqnarray}\label{RHO-AiBj}
\rho=\sum_{n=1}^{d_A^2}\sum_{m=1}^{d_B^2} r_{nm} A_n\otimes B_m.
\end{eqnarray}
In this context, the coefficients \(r_{nm}\) represent the elements constituting the \(d_A^2 \times d_B^2\) correlation matrix denoted as \(R\). With the aid of orthogonal matrices \(U\) (a \(d_A^2 \times d_A^2\) matrix) and \(V\) (a \(d_B^2 \times d_B^2\) matrix), one can derive the singular value decomposition of \(R\) such that \(R=U{\text{diag}}(c_1,\dots,c_L,0,\dots,0)V^T\), where \(\{c_k: k=1,2,\dots,L\}\) are the nonzero singular values and \(L=\text{rank}(R)\) signifies the rank of the correlation matrix. Moreover, one can introduce the matrices \(S_n=\sum_{n'}u_{nn'}A_{n'}\) and \(F_{m}=\sum_{m'}v_{mm'}B_{m'}\), where \(u_{nn'}\) and \(v_{mm'}\) are matrix elements of \(U\) and \(V\), respectively. Then, the state \(\rho\) given by Eq.~\eqref{RHO-AiBj} can be written such that \cite{Dakic2020NonzeroQDCondition}
\begin{eqnarray}\label{RHO-SkFk}
\rho=\sum_{k=1}^{L} c_k S_k \otimes F_k.
\end{eqnarray}
Of significance, the state \(\rho\) defined in Eq.~\eqref{RHO-SkFk} exhibits two characteristics: first, \(L\) determines how many product operators are needed to represent \(\rho\) \cite{Gessner2012LocalCreationofQD}; second, the commutation relations of the local operators \(S_k\) (\(F_k\)) determine the local quantumness of the state in the subsystem \(\mathcal{H}_A\) (\(\mathcal{H}_B\)) \cite{Dakic2020NonzeroQDCondition}, a topic elaborated upon in Section \ref{Subsec:Set-of-Nonzero-QD}.


\subsection{Quantum discord}\label{Subsec:QuantumDiscord}

Delving into the non-classical characteristics of a particular state stands as a key focus in the domain of quantum information theory. In this context, the pivotal role of QD \cite{Zurek2001-QD, Henderson2001-QCC} lies in its capacity to illuminate the non-classical correlations that are inherently present in bipartite quantum systems. Mathematically, QD (\(\mathcal{D}\)) \cite{Zurek2001-QD} is delineated as the disparity between the total correlation (\(\mathcal{J}\)) and the classical correlation (\(\mathcal{C}\)) \cite{Henderson2001-QCC}, thus expressed as \(\mathcal{D} = \mathcal{J} - \mathcal{C}\).

The total correlation \(\mathcal{J}\) of a bipartite quantum state \(\rho_{AB}\) can be quantified by the quantum mutual information. It is defined as the difference between the von Neumann entropy of the joint system \(\rho_{AB}\) and the sum of the von Neumann entropies of the subsystems \(\rho_A\) and \(\rho_B\), that is,
\begin{eqnarray}\label{Def:TotalCorr}
\mathcal{J} = S(\rho_{A}) + S(\rho_{B}) - S(\rho_{AB}).
\end{eqnarray}
Here, \(S(\rho_A)\) (\(S(\rho_B)\)) and \(S(\rho_{AB})\) are the von Neumann entropies of the subsystem \(\rho_A\) (\(\rho_B\)) and joint system \(\rho_{AB}\), respectively. On the other hand, the classical correlation \(\mathcal{C}\) is determined by optimizing the measurement basis \(\{ \Pi_{A}^{a} \}\) on subsystem \(A\), and subsequently calculating the (quantum) mutual information for the resulting post-measurement states \(\{ \rho_{B|a} \}\). The classical correlation \(\mathcal{C}\) is then defined as the maximum (classical) mutual information over all possible measurements:
\begin{eqnarray}\label{Def:ClassicalCorr}
\mathcal{C} = \max_{\{ \Pi_{A}^{a} \}} \left( S(\rho_{B}) - S(\rho_{B|\Pi_{A}^{a}}) \right),
\end{eqnarray}
where
\begin{eqnarray}
S(\rho_{B|\Pi_{A}^{a}}) = \sum_{a} {p_{a}} \, S(\rho_{B|a}).
\end{eqnarray}
Here, \(\{p_{a}\geq 0\}\) denotes the probability of obtaining outcome \(a\) after the measurement \(\{ \Pi_{A}^{a} \}\) on subsystem \(A\). More precisely, one has
\begin{eqnarray}
\left\{p_a = \text{Tr}\left({\Pi_{A}^{a}}^{\dagger} \Pi_{A}^{a} \rho_{AB}\right) ; \,\rho_{B|a} = \frac{\text{Tr}_A\Big(\Pi_{A}^{a} \rho_{AB} \Pi_{A}^{a}\Big)}{p_a}\right\}
\end{eqnarray}
The difference between the total correlation \(\mathcal{J}\) given by Eq.~\eqref{Def:TotalCorr} and the classical correlation \(\mathcal{C}\) given by Eq.~\eqref{Def:ClassicalCorr} encapsulates the QD \cite{Zurek2001-QD, Henderson2001-QCC}, that is,
\begin{eqnarray}\label{Def:QD}
\mathcal{D} &=& \mathcal{J} - \mathcal{C} \nonumber \\
&=& S(\rho_A)-S(\rho_{AB})+\underset{\{\Pi_{A}^{a}\}}{\text{min}}\sum_a {p_{a}} \, S(\rho_{B|a}).
\end{eqnarray}
Note that, QD defined in Eq.~\eqref{Def:QD} has been only calculated for some special states due to the complexity of the minimization problem \cite{GAdesso2017-BookQD}, and it still needs to be developed. In addition, another measure of QD, namely the geometric measure \cite{Dakic2020NonzeroQDCondition}, is employed to address the challenge posed by this minimization. The geometric measure of QD was introduced as \cite{Dakic2020NonzeroQDCondition}
\begin{eqnarray}\label{GeometricDiscord}
D_{G}(\rho) = \underset{\chi \in \Omega_0}{\text{min}} \|\rho - \chi\|^2 = \underset{\chi \in \Omega_0}{\text{min}} \text{Tr}\left[(\rho - \chi)^2\right],
\end{eqnarray}
where \(\Omega_0\) denotes the set of zero-discord states. In this study, one of our aims is to deepen our understanding of QD by explicitly demonstrating how it changes under local quantum operations.


\subsection{Set of nonzero discord states}\label{Subsec:Set-of-Nonzero-QD}

In the discussion below, the focus is on the necessary and sufficient condition for the existence of (non)zero quantum discord. Note that, although the criterion is explained with respect to the first subsystem, the same is also valid for the second subsystem. Now, let us consider a bipartite state \(\rho\) in dimension \(d\). It is of zero discord state with respect to the first subsystem if and only if there exists a von Neumann measurement \(\{\Pi_i=\ket{\psi_i}\bra{\psi_i}\}\) \cite{Dakic2020NonzeroQDCondition} such that
\begin{eqnarray}\label{ZeroDiscord-Pii}
\sum_{i}\big(\Pi_i\otimes {\dsone}_B\big) \rho \big(\Pi_i\otimes {\dsone}_B\big)=\rho.
\end{eqnarray}
One then concludes from Eq.~\eqref{ZeroDiscord-Pii} that a zero-discord state (with respect to the first subsystem) is of the form
\begin{eqnarray}\label{ZeroDiscord-PsiPsi}
\rho=\sum_{i}p_i\ket{\psi_i}\bra{\psi_i}\otimes\rho_B^{(i)}, \quad \Big(p_i \geq 0; \ \sum_{i}p_i=1\Big),
\end{eqnarray}
where \(\{\ket{\psi_i}\}\) is some orthonormal basis set for the first subsystem and \(\{\rho_B^{(i)}\}\) are the quantum states for the second subsystem. As mentioned earlier, this consideration can also be done with respect to the second subsystem, that is, a state \(\rho=\sum_{i}p_i \rho_A^{(i)} \otimes \ket{\phi_i}\bra{\phi_i}\) is a zero-discord state (with respect to the second subsystem), where \(\{\ket{\phi_i}\}\) is some orthonormal basis set for the second subsystem and \(\{\rho_A^{(i)}\}\) are the quantum states for the first subsystem.

As outlined in Ref.~\cite{Dakic2020NonzeroQDCondition}, a more convenient form of the necessary and sufficient condition is also possible. Given the state \(\rho\) as specified in Eq.~\eqref{RHO-SkFk}, Eq.~\eqref{ZeroDiscord-Pii} assumes the following form:
\begin{eqnarray}\label{NecSuffCond-Sk}
\sum_{k=1}^{L}c_k \Big(\sum_{i} \Pi_i S_k \Pi_i\Big) \otimes F_k = \sum_{k=1}^{L} c_k S_k \otimes F_k,
\end{eqnarray}
which is equivalent to the set of conditions \(\{[S_k, \Pi_i]=0: k=1, 2, \dots, L\}\) for all \(i\). This means that the set of operators \(\{S_k: k=1,2,\dots,L\}\) has a common eigenbasis defined by the set of projectors \(\{\Pi_i\}\). Therefore, the set \(\{\Pi_i\}\) exists if and only if \cite{Dakic2020NonzeroQDCondition}
\begin{eqnarray}\label{ZeroDiscord-SmuSnuComu}
[S_{\mu}, S_{\nu}]=0, \quad \mu, \nu = 1, 2, \dots, L,
\end{eqnarray}
where \(L=\text{R}\leq\text{min}\{d_A^2,d_B^2\}\). Here it is necessary to verify at most \(L(L-1)/2\) commutators in Eq.~\eqref{ZeroDiscord-SmuSnuComu} to ascertain the absence of discord in the given state. Moreover, going back to the state of zero discord given by Eq.~\eqref{ZeroDiscord-PsiPsi}, one can set \(i=1,2,\dots,d_A\) which bounds \(R\) to \(L\leq d_A\). Then, the rank of the correlation matrix \(R\) is a simple discord witness: if \(L>d_A\), the state \(\rho\) has a nonzero discord \cite{Dakic2020NonzeroQDCondition}.


\section{Generating Quantum Dissonance}\label{Sec:Results} 

In this section, we explore the conversion of initially classically correlated states, which exhibit zero discord, into states with non-zero discord through local quantum operations. Our study centers on the analysis of (two-qubit) Werner state as target state(s).


\subsection{Werner states}\label{Subsec:WernerStates}

\begin{figure}[htbp!]
\vspace*{13pt}
\centerline{\epsfig{file=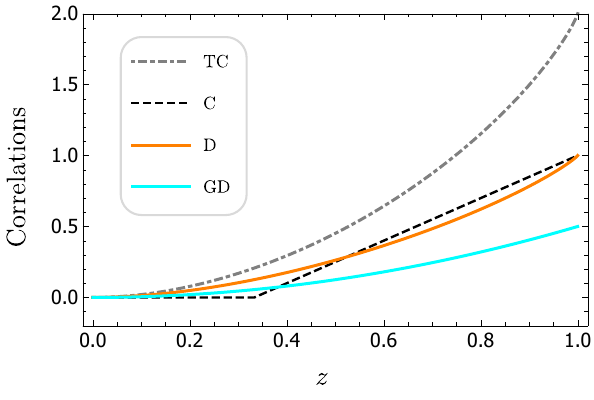, width=9.2cm}} 
\vspace*{13pt}
\fcaption{\label{Fig:Werner}The graph shows the values of the total correlation (TC, gray dotted-dashed line), the concurrence (C, black dashed line), the discord (D, orange solid line), and the geometric discord (GD, cyan solid line) for two-dimensional Werner state \(\rho_{\text{wer}}(z)\) given in Eq.~\eqref{Werner-z}. For \(z\in (\frac{1}{3},1]\) the state is entangled; however, for \(z\in [0,\frac{1}{3}]\) it is a separable but a nonzero discord state.}
\end{figure}

The two-qubit Werner state is given by the following density operator
\begin{eqnarray}\label{Werner-z}
\rho_{\text{wer}}(z)=z\ket{\Psi^{-}}\bra{\Psi^{-}}+(\frac{1-z}{4}) {\dsone}_2 \otimes {\dsone}_2, \quad z\in [0,1].
\end{eqnarray}
Here, \({\dsone}_2\) denotes the two-dimensional identity operator and \(\ket{\Psi^{-}}=(\ket{01}-\ket{10})/\sqrt{2}\) is the singlet state. The Werner state \(\rho_{\text{wer}}(z)\) given in Eq.~\eqref{Werner-z} is characterized by a parameter \(z\in [0,1]\). Namely, for \(z\in (\frac{1}{3},1]\) the Werner state \(\rho_{\text{wer}}(z)\) is entangled; however, for \(z\in [0,\frac{1}{3}]\) the Werner state \(\rho_{\text{wer}}(z)\) is separable. To show that \(\rho_{\text{wer}}(z)\) is separable in the range \(z\in [0,\frac{1}{3}]\), one can use the positive partial transpose criterion \cite{Horodecki1996-PPT} or concurrence \cite{Wootters1997-Werner}, which are two well known entanglement measures for \(2 \times 2\) quantum states. Importantly, for \(z\in (0,1]\) the Werner state \(\rho_{\text{wer}}(z)\) has a nonzero discord. In other words, although entanglement of \(\rho_{\text{wer}}(z)\) vanishes in the range \(z\in (0,\frac{1}{3}]\), it has a nonzero discord. Figure \ref{Fig:Werner} clearly shows relevant correlations for the Werner state \(\rho_{\text{wer}}(z)\).

As we mentioned before, our aim is to obtain the separable Werner states (which have nonzero discord) from a classically correlated state by using local quantum operations. To this end, we need to express \(\rho_{\text{wer}}(z)\) in terms of product states. We start with the following convex combination of product states for the separable Werner state as in Refs.~\cite{Wootters1997-Werner,Azuma2006-Werner}:
\begin{eqnarray}\label{Werner-SEP1}
\rho_{\text{wer}}(z)=\sum_{j=0}^{3} \ket{\eta_j}\bra{\eta_j}, \quad z\in [0,\frac{1}{3}].
\end{eqnarray}
The states \(\{\ket{\eta_j}: j=0,1,2,3\}\) appear in Eq.~\eqref{Werner-SEP1} are given by
\begin{gather}\begin{aligned}\label{Eta-States}
\ket{\eta_0}&=\frac{1}{2}\left(e^{i\theta_1}\ket{x_1}+e^{i\theta_2}\ket{x_2}+e^{i\theta_3}\ket{x_3}+e^{i\theta_4}\ket{x_4}\right),\\
\ket{\eta_1}&=\frac{1}{2}\left(e^{i\theta_1}\ket{x_1}+e^{i\theta_2}\ket{x_2}-e^{i\theta_3}\ket{x_3}-e^{i\theta_4}\ket{x_4}\right),\\
\ket{\eta_2}&=\frac{1}{2}\left(e^{i\theta_1}\ket{x_1}-e^{i\theta_2}\ket{x_2}+e^{i\theta_3}\ket{x_3}-e^{i\theta_4}\ket{x_4}\right),\\
\ket{\eta_3}&=\frac{1}{2}\left(e^{i\theta_1}\ket{x_1}-e^{i\theta_2}\ket{x_2}-e^{i\theta_3}\ket{x_3}+e^{i\theta_4}\ket{x_4}\right),
\end{aligned}\end{gather}
with \(\ket{x_1}=\frac{\sqrt{1+3z}}{2i}\ket{\Psi^-}\), \(\ket{x_2}=\frac{\sqrt{1-z}}{2}\ket{\Psi^+}\),
\(\ket{x_3}=\frac{\sqrt{1-z}}{2}\ket{\Phi^-}\), \(\ket{x_4}=\frac{\sqrt{1-z}}{2i}\ket{\Phi^+}\) \cite{Wootters1997-Werner,Azuma2006-Werner}.
Note that \(\ket{\Psi^{\pm}}=(\ket{01}\pm\ket{10})/{\sqrt{2}}\) and \(\ket{\Phi^{\pm}}=(\ket{00}\pm\ket{11})/{\sqrt{2}}\) are Bell states.
Also, it is straightforward to show that \(\braket{\eta_j}{\eta_k}=z/4\) for \(j,k=0, 1, 2, 3\)  (\(j\neq k\)) and \(\braket{\eta_j}{\eta_j}=1/4\).
The parameters \(\{\theta_l: l=1,2,3,4\}\) appear in Eq.~\eqref{Eta-States} satisfy the following equation
\begin{eqnarray}\label{theta}
e^{-2i\theta_1}(1+3z)+\left(e^{-2i\theta_2}+e^{-2i\theta_3}+e^{-2i\theta_4}\right)(1-z)=0.
\end{eqnarray}
Without loss of generality, two of phases can be chosen as \(\theta_1=0\) and \(\theta_2=\pi/2\).
By choosing such that, from Eq.~\eqref{theta} one can get \(\cos\theta_3=\sqrt{\frac{1-3z}{2(1-z)}}\), \(\sin{\theta_3}=\sin{\theta_4}=\sqrt{\frac{1+z}{2(1-z)}}\),
and \(\cos\theta_4=-\sqrt{\frac{1-3z}{2(1-z)}}\). With all this information at hand, we are now ready to find the product states \(\{\ket{\eta_j}: j=0,1,2,3\}\) for \(z \in [0,\frac{1}{3}]\). In what follows, we explicitly discuss how one can obtain the product state \(\ket{\eta_0}\) for \(z=\frac{1}{3}\). Combining \(\{\ket{x_l}: l=1,2,3,4\}\) and Eq.~\eqref{theta} with \(\ket{\eta_0}\) given in Eq.~\eqref{Eta-States}, the product state \(\ket{\eta_0}\) can be written such that
\begin{eqnarray}\begin{aligned}\label{Eta0-explicit}
\ket{\eta_0}&=\frac{1}{2}\big(\ket{x_1}+i\ket{x_2}+i\ket{x_3}+i\ket{x_4}\big) \\
&=\frac{1}{2}\Big(-\frac{i}{\sqrt{2}}\ket{\Psi^-}+\frac{i}{\sqrt{6}}\ket{\Psi^+}
+\frac{i}{\sqrt{6}}\ket{\Phi^-}+\frac{1}{\sqrt{6}}\ket{\Phi^+}\Big) \\
&=\frac{1}{2}\big(\alpha_{0,0}\ket{0}+\alpha_{0,1}\ket{1}\big)\otimes\big(\beta_{0,0}\ket{0}+\beta_{0,1}\ket{1}\big) \\
&=\frac{1}{2}\ket{\Psi_0}\otimes \ket{\Phi_0}.
\end{aligned}\end{eqnarray}
Here, two states \(\ket{\Psi_0}\) and \(\ket{\Phi_0}\) are normalized, that is, \(|\alpha_{0,0}|^2+|\alpha_{0,1}|^2=1\) and \(|\beta_{0,0}|^2+|\beta_{0,1}|^2=1\). After elementary calculations, one can conveniently obtain coefficients \(\alpha_{0,i}\) and \(\beta_{0,i}\) (\(i=0,1\)). We hence find
\begin{eqnarray}\label{Product-ETA0}
\ket{\Psi_0} &=& \kappa{e^{i\frac{\pi}{2}}}\left[\big(1-\sqrt{3}\big)\ket{0}-\big({1+i}\big)\ket{1}\right], \nonumber \\
\ket{\Phi_0} &=& \kappa\left[\big({i-1}\big)\ket{0}+\big(\sqrt{3}-1\big)\ket{1}\right],
\end{eqnarray}
where \(\kappa \equiv \sqrt{\frac{3+\sqrt{3}}{12}}\).
Similarly, other states can also be written in terms of product states:
\begin{eqnarray}\begin{aligned}
\ket{\eta_j}&=\frac{1}{2}\ket{\Psi_j}\otimes\ket{\Phi_j} \\
&=\frac{1}{2}\big(\alpha_{j,0}\ket{0}+\alpha_{j,1}\ket{1}\big)\otimes\big(\beta_{j,0}\ket{0}+\beta_{j,1}\ket{1}\big),
\end{aligned}\end{eqnarray}
such that \(|\alpha_{j,0}|^2+|\alpha_{j,1}|^2=1\) and \(|\beta_{j,0}|^2+|\beta_{j,1}|^2=1\) for \(j=1,2,3\). Following the same steps as with \(\ket{\eta_0}\) in Eq.~\eqref{Eta0-explicit}, the pairs \(\{\ket{\Psi_j}; \ket{\Phi_j}\}\) are found to be
\begin{eqnarray}\label{Product-ETA1}
\ket{\Psi_1}=\kappa{e^{i\frac{\pi}{2}}}\left[\big(1-\sqrt{3}\big)\ket{0}+\big({1+i}\big)\ket{1}\right], \quad
\ket{\Phi_1}=\kappa\left[\big({1-i}\big)\ket{0}+\big(\sqrt{3}-1\big)\ket{1}\right];
\end{eqnarray}
\begin{eqnarray}\label{Product-ETA2}
\ket{\Psi_2}={\bar{\kappa}}{e^{-i\frac{\pi}{2}}}\left[\big(\sqrt{3}+1\big)\ket{0}+\big({1-i}\big)\ket{1}\right], \quad
\ket{\Phi_2}={\bar{\kappa}}\left[-\big({1+i}\big)\ket{0}+\big(\sqrt{3}+1\big)\ket{1}\right];
\end{eqnarray}
\begin{eqnarray}\label{Product-ETA3}
\ket{\Psi_3}={\bar{\kappa}}{e^{-i\frac{\pi}{2}}}\left[\big(\sqrt{3}+1\big)\ket{0}+\big({i-1}\big)\ket{1}\right], \quad
\ket{\Phi_3}={\bar{\kappa}}\left[\big({1+i}\big)\ket{0}+\big(\sqrt{3}+1\big)\ket{1}\right],
\end{eqnarray}
where \(\bar{\kappa} \equiv \sqrt{\frac{3-\sqrt{3}}{12}}\).
From Eqs.~\eqref{Product-ETA0}, \eqref{Product-ETA1}, \eqref{Product-ETA2}, and \eqref{Product-ETA3}, it is easy to show that \(\braket{\Psi_j}{\Phi_j}=0\), \(\braket{\Psi_j}{\Psi_k}\neq 0\), \(\braket{\Phi_j}{\Phi_k}\neq 0\), and \(\braket{\Psi_j}{\Psi_k}\braket{\Phi_j}{\Phi_k}=\frac{1}{3}\) for \(j,k=0,1,2,3\) such that \(j \neq k\).
All in all, we rewrite the separable Werner state for \(z=\frac{1}{3}\) such that
\begin{eqnarray}\label{product-W-z1/3}
\rho_{\text{wer}}({1}/{3})=\frac{1}{4}\sum_{j=0}^{3}\ket{\Psi_j}\bra{\Psi_j}\otimes\ket{\Phi_j}\bra{\Phi_j},
\end{eqnarray}
where the states \(\ket{\Psi_j}\) and \(\ket{\Phi_j}\) are given in Eqs.~\eqref{Product-ETA0}, \eqref{Product-ETA1}, \eqref{Product-ETA2}, and \eqref{Product-ETA3}. Even though we clearly demonstrate the calculations here for the value of \(z\) equal to \(\frac{1}{3}\), the consideration here applies to any value of \(z\) within the range \([0,\frac{1}{3}]\). Thus, we demonstrate how to obtain the product forms of Werner states \cite{Wootters1997-Werner,Azuma2006-Werner} and the states included in this product forms.

We conclude this section with another important piece of information provided by Eq.~\eqref{product-W-z1/3}. That is, we ensure that rank of the correlation matrix of the state that we aim to obtain by local operations, separable Werner state for \(z=\frac{1}{3}\), is equal to \(4\). Then, to obtain the target state \(\rho_{\text{wer}}(z)\) for \(z \in (0,\frac{1}{3}]\), we initially must have a rank \(4\) (or greater than \(4\)) density matrix.


\subsection{Generating Werner states via local operations}\label{Subsec:GeneratingNonzeroWerner}

\begin{figure} [htbp]
\vspace*{13pt}
\centerline{\epsfig{file=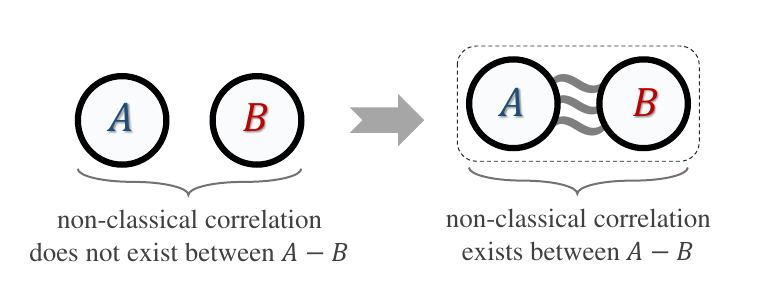, width=9.2cm}} 
\vspace*{13pt}
\fcaption{\label{figure1}Two parties, Alice \((A)\) and Bob \((B)\), share a state, where Alice and Bob have \(k\)-qubits \(\{\mathcal{A}_1, \dots, \mathcal{A}_k\}\) and \(\{\mathcal{B}_1, \dots, \mathcal{B}_k\}\), respectively, in a way that the qubits \(\mathcal{A}_j\) and \(\mathcal{B}_j\) (\(j=1, \dots, k\) where \(k=2,3\)) are classically correlated. Following our proposed method results in the generation of non-classical correlation between \(A\) and \(B\). The explicit quantum operations used in the task are described in the paper.}
\end{figure}

Having introduced the way to express separable Werner states in terms of product states \cite{Wootters1997-Werner,Azuma2006-Werner}, we now consider the following scenario. Suppose we are given an initial state shared between Alice (\(A\)) and Bob (\(B\)), where \(A\) consists of qubits \(\{\mathcal{A}_1, \dots, \mathcal{A}_k\}\) and \(B\) consists of qubits \(\{\mathcal{B}_1, \dots, \mathcal{B}_k\}\). They are prepared locally where the qubits \(\mathcal{A}_j\) and \(\mathcal{B}_j\) (\(j=1, \dots, k\)) are classically correlated. Our main goal here is to demonstrate whether non-classical correlation --- quantum dissonance in this case --- can be generated between \(A\) and \(B\) via local operations. Figure \ref{figure1} illustrates this consideration.

We examine this problem taking \(k=2, 3\). Let us first consider the case \(k=2\), where the initial state is given by
\begin{eqnarray}\label{Class-Corr-k2}
\rho_{AB}
&=&\bigotimes_{j=1}^{2}\frac{1}{2}\left(\sum_{i=0}^{1}\ket{i}\bra{i}_{\mathcal{A}_j}\otimes\ket{i}\bra{i}_{\mathcal{B}_j}\right).
\end{eqnarray}
Here in Eq.~\eqref{Class-Corr-k2} both subsystems consist of two qubits such that \(\mathcal{A}_j\) and \(\mathcal{B}_j\) are classically correlated; but, the pairs \((\mathcal{A}_1, \mathcal{A}_2)\) and \((\mathcal{B}_1, \mathcal{B}_2)\) are uncorrelated. Clearly, the rank of the state described in Eq.~\eqref{Class-Corr-k2} is 4, as it should be (see section \ref{Subsec:CorrelationRank}).

In section \ref{Subsec:WernerStates} we provided detailed explanations for \(z=\frac{1}{3}\) and mentioned that same calculations would be done for all values of \(z\) within the range \([0, \frac{1}{3}]\). Here, we also consider the case where \(z\) equals \(\frac{1}{3}\). Thus, we employ the states provided in Eqs.~\eqref{Product-ETA0}, \eqref{Product-ETA1}, \eqref{Product-ETA2}, and \eqref{Product-ETA3} when defining local operations. In this regard, we construct the local operations for the subsystem \(A = \mathcal{A}_1 \otimes \mathcal{A}_2\) (or equivalently \(\mathcal{A}_1\mathcal{A}_2\)) such that
\begin{eqnarray}\label{M-A-twoqubits}
\left\{M_A^i\right\}_{i=0}^3= \Big\{\ket{0\Psi_0}\bra{00}_{\mathcal{A}_1\mathcal{A}_2}, \, \ket{0\Psi_1}\bra{01}_{\mathcal{A}_1\mathcal{A}_2}, \, \ket{1\Psi_2}\bra{10}_{\mathcal{A}_1\mathcal{A}_2}, \, \ket{1\Psi_3}\bra{11}_{\mathcal{A}_1\mathcal{A}_2}\Big\},
\end{eqnarray}
where \(\sum_{i=0}^{3}\left(M_A^i\right)^{\dag}M_A^i = {\dsone}_4 \), and the local operations for the subsystem \(B = \mathcal{B}_1 \otimes \mathcal{B}_2\) (or equivalently \(\mathcal{B}_1\mathcal{B}_2\)) such that
\begin{eqnarray}\label{M-B-twoqubits}
\left\{M_B^i\right\}_{i=0}^3= \Big\{\ket{0\Phi_0}\bra{00}_{\mathcal{B}_1\mathcal{B}_2}, \, \ket{0\Phi_1}\bra{01}_{\mathcal{B}_1\mathcal{B}_2}, \, \ket{1\Phi_2}\bra{10}_{\mathcal{B}_1\mathcal{B}_2}, \, \ket{1\Phi_3}\bra{11}_{\mathcal{B}_1\mathcal{B}_2}\Big\},
\end{eqnarray}
where one can easily check that \(\sum_{i=0}^{3}(M_B^i)^{\dag}M_B^i = {\dsone}_4 \) indeed. Here, needless to say, \({\dsone}_4\) is the \(4 \times 4\) is identity matrix. By performing the constructed local operations as given in Eqs.~\eqref{M-A-twoqubits} and \eqref{M-B-twoqubits} on subsystems \(A\) and \(B\) sequentially, the state \(\rho_{AB}\) given by Eq.~\eqref{Class-Corr-k2} is transformed into state \(\rho_{AB}'\), that is,
\begin{eqnarray}\label{rhotilde-4qubits}
{\rho}_{AB}' &=& \sum_{i=0}^{3} \Big(M_A^i \otimes M_B^i\Big) \rho_{AB} \Big(\big(M_A^i\big)^{\dagger} \otimes (M_B^i)^{\dagger}\Big).
\end{eqnarray}
Taking partial trace of the qubits \(\mathcal{A}_1\) and \(\mathcal{B}_1\) of the state
\({\rho}_{AB}'\) in Eq. \eqref{rhotilde-4qubits}
gives the target Werner state:
\begin{eqnarray}
{\text{Tr}}_{\mathcal{A}_1 \mathcal{B}_1}[{\rho}_{AB}']=\frac{1}{4}\sum_{i=0}^{3}\ket{\Psi_i}\bra{\Psi_i}_{\mathcal{A}_2} \otimes \ket{\Phi_i}\bra{\Phi_i}_{\mathcal{B}_2}
= \rho_{\text{wer}}({1}/{3}).
\end{eqnarray}
Consequently, by employing local operations in the form of \eqref{M-A-twoqubits} and \eqref{M-B-twoqubits}, it is possible to transform a classically correlated state, initially possessing zero discord, into Werner states characterized by non-classical correlations exhibiting discord values greater than zero.

We now look into the scenario with \(k=3\), that is, two subsystems denoted as \(A\) and \(B\) comprise three qubits each. Specifically, qubits \(({\mathcal{A}_1}, {\mathcal{B}_1})\), \(({\mathcal{A}_2}, {\mathcal{B}_2})\), and \(({\mathcal{A}_3}, {\mathcal{B}_3})\) are classically correlated. The state of the total system is then of the form
\begin{eqnarray}\label{ClassCorr-k3}
\rho_{AB} = \rho_{\mathcal{A}_1 \mathcal{B}_1}\otimes\rho_{\mathcal{A}_2 \mathcal{B}_2}\otimes\rho_{\mathcal{A}_3 \mathcal{B}_3},
\end{eqnarray}
where
\begin{eqnarray}
\rho_{\mathcal{A}_j \mathcal{B}_j} = \frac{1}{2} \sum_{i=0}^{1}\ket{i}\bra{i}_{\mathcal{A}_j} \otimes\ket{i}\bra{i}_{\mathcal{B}_j}, \quad j = 1, 2, 3.
\end{eqnarray}
We adopt an approach that involves the systematic incorporation of unitary operations. We begin with the unitary transformation performed by Alice (\(A\)) which is expressed as follows:
\begin{eqnarray}\label{LocalUni-Alice}
U_A &\equiv&  U_{{\mathcal{A}_1} {\mathcal{A}_2} {\mathcal{A}_3}} \nonumber \\
&=& \sum_{m,n=0}^{1}\bigg[\ket{m}\bra{m}_{\mathcal{A}_1} \otimes \ket{n}\bra{n}_{\mathcal{A}_2} \otimes \Big(\ket{\Psi_{2m+n}}\bra{0}_{\mathcal{A}_3}+\ket{\Phi_{2m+n}}\bra{1}_{\mathcal{A}_3}\Big)\bigg].
\end{eqnarray}
The unitarity condition on \(U_A\), \(U_{A}^{\ast} U_{A} = U_{A} U_{A}^{\ast} = {\dsone}\), imposes the constraints \(\braket{\Psi_l}{\Phi_l}=0\) \((l=0,1,2,3)\), which are corroborated by Eqs.~\eqref{Product-ETA0}, \eqref{Product-ETA1}, \eqref{Product-ETA2}, and \eqref{Product-ETA3}. Similarly, the unitary transformation performed by Bob (\(B\)) is expressed as follows:
\begin{eqnarray}\label{LocalUni-Bob}
U_B &\equiv&  U_{{\mathcal{B}_1} {\mathcal{B}_2} {\mathcal{B}_3}} \nonumber \\
&=& \sum_{m,n=0}^{1} \bigg[\ket{m}\bra{m}_{\mathcal{B}_1} \otimes \ket{n}\bra{n}_{\mathcal{B}_2} \otimes \Big(\ket{\Phi_{2m+n}}\bra{0}_{\mathcal{B}_3}+\ket{\Psi_{2m+n}}\bra{1}_{\mathcal{B}_3}\Big)\bigg],
\end{eqnarray}
where the unitarity condition on \(U_B\), \(U_{B}^{\ast} U_{B} = U_{B} U_{B}^{\ast} = {\dsone}\), imposes identical constraints as those on \(U_A\). After applying the unitary transformations performed by Alice and Bob to their respective systems, the resulting state is
\begin{eqnarray}\label{QDmethod3}
\tilde{\rho}_{AB} &=& \Big(U_{A} \otimes U_{B}\Big) \rho_{AB} \Big(U^{\dag}_{A} \otimes U^{\dag}_{B}\Big) \nonumber \\
&=& \frac{1}{8} \sum_{m,n=0}^{1} \Big[\ket{mn}\bra{mn}_{\mathcal{A}_1\mathcal{A}_2} \ket{\Psi_{2m+n}}\bra{\Psi_{2m+n}}_{\mathcal{A}_3} \otimes \ket{mn}\bra{mn}_{\mathcal{B}_1\mathcal{B}_2} \ket{\Phi_{2m+n}}\bra{\Phi_{2m+n}}_{\mathcal{B}_3} \nonumber \\
& & + \ket{mn}\bra{mn}_{\mathcal{A}_1\mathcal{A}_2} \ket{\Phi_{2m+n}}\bra{\Phi_{2m+n}}_{\mathcal{A}_3} \otimes \ket{mn}\bra{mn}_{\mathcal{B}_1\mathcal{B}_2} \ket{\Psi_{2m+n}}\bra{\Psi_{2m+n}}_{\mathcal{B}_3} \Big].
\end{eqnarray}
In the final step, we execute the trace operation on the qubits labeled \({\mathcal{A}_1}\), \({\mathcal{A}_2}\), \({\mathcal{B}_1}\), and \({\mathcal{B}_2}\) within the state \(\tilde{\rho}_{AB}\) as described in Eq.~\eqref{QDmethod3}. This operation yields
\begin{eqnarray}
\tilde{\rho}_{\mathcal{A}_3\mathcal{B}_3}
&=& \text{Tr}_{{\mathcal{A}_1\mathcal{A}_2}{\mathcal{B}_1\mathcal{B}_2}} \left[\tilde{\rho}_{AB}\right] \nonumber \\
&=& \frac{1}{2}\left(\frac{1}{4}\sum_{i=0}^{3} \ket{\Psi_i}\bra{\Psi_i}_{\mathcal{A}_3} \otimes \ket{\Phi_i}\bra{\Phi_i}_{\mathcal{B}_3}
+\frac{1}{4}\sum_{i=0}^{3}\ket{\Phi_i}\bra{\Phi_i}_{\mathcal{A}_3} \otimes \ket{\Psi_i}\bra{\Psi_i}_{\mathcal{B}_3} \right) \nonumber \\
&=& \frac{1}{2}\Big(\rho_{\text{wer}}({1}/{3}) + \rho_{\text{wer}}({1}/{3})\Big) \nonumber \\
&=& \rho_{\text{wer}}({1}/{3}).
\end{eqnarray}
As a result, by applying the unitary operations described in Eqs.~\eqref{LocalUni-Alice} and \eqref{LocalUni-Bob}, it is possible to convert a classically correlated state given by Eq.~\eqref{ClassCorr-k3}, originally without discord, into Werner states, exhibiting nonzero QD (i.e., quantum dissonance) as a form of non-classical correlation.


\section{Conclusion}\label{Sec:Conclusion}

Quantum state transformations offer fundamental insights into how quantum systems evolve and can be manipulated, with wide-ranging applications in quantum information processing. In this paper, we have studied the state transformations involved in non-classical correlations, focusing particularly on QD as a measure to quantify such correlations beyond entanglement. We have presented two procedures for generating separable Werner states with nonzero QD, commonly referred to as quantum dissonance. These methods provide a systematic approach that differs from existing techniques by considering the inclusion of local quantum operators. Our study, involving explicit forms of the states of the subsystems constituting Werner states, could enhance comprehension of the interplay between QD and nonorthogonality.

Understanding the role and impact of nonorthogonality in QD is crucial indeed and can help us to successfully address related issues. For instance, investigating whether certain classes of nonorthogonal states inherently possess higher/lower discord could shed light on the detailed nature of QCs. Additionally, exploring potential links between coherence \cite{Baumgratz2014RTCoherence} (based on orthogonal basis states), superposition \cite{Theurer2017RTSuperposition, Torun2021RTSuperposition, Senyasa2022GoldenSuperposition, Torun2023GoldenSuperposition, Pusuluk2024Superposition} (where basis states are not necessarily orthogonal), and QD constitutes a significant contribution. Searching for answers to such questions not only enriches understanding of QD but also may facilitate the derivation of new measure(s) for superposition as a resource, given the challenge of introducing easily applicable quantifier due to the existence of nonorthogonality. This serves as motivation for our study, and we believe it contributes significantly.


\nonumsection{References}
\noindent


\end{document}